\documentclass[10pt,conference]{IEEEtran}
\IEEEoverridecommandlockouts

\usepackage[dvipsnames]{xcolor}

\usepackage{cite}
\usepackage{amsmath,amssymb,amsfonts}
\usepackage{algorithm}
\usepackage{algpseudocode} 

\usepackage{graphicx}
\usepackage{textcomp}
\usepackage{xcolor}

\usepackage{kotex}
\usepackage{graphicx}
\usepackage{hyperref}
\usepackage{adjustbox}
\usepackage{booktabs}
\usepackage{multirow}
\usepackage{tabularx}
\usepackage{siunitx}
\usepackage{verbatim}
\usepackage{ulem}

\usepackage{xcolor}
\usepackage{verbatim}
\usepackage{array}

\usepackage{amsmath} 

\usepackage{algpseudocode}  

\usepackage{listings}
\usepackage{xcolor}
\usepackage{courier}  
\usepackage{textcomp} 
\definecolor{placeholderpurple}{HTML}{A02B93}
\definecolor{lightgreen}{RGB}{200,255,200}
\definecolor{lightyellow}{RGB}{255,255,200}

\lstdefinestyle{textstyle}{
    basicstyle=\ttfamily\tiny,
    stringstyle=\color{black},
    numbers=none,              
    breaklines=true,
    captionpos=b,
    frame=lines,
    backgroundcolor=\color{white},
    moredelim=[s][\color{placeholderpurple}]{\{}{\}}, 
    escapeinside={(*@}{@*)} 
}

\lstdefinelanguage{ALPG}{
    keywords={JSR, NOP, STPS, MODULE, START, JNI2, CYP1, CYP2, CYP3, CYP4, CMDI, RPRE, TMRSI, DBMAP},
    keywordstyle=\color{blue}\bfseries,
    ndkeywords={CE0, TS1, TS2, TS6, TS7, R, G_LF001_CMDI, G_LF001_ADD, G_LF001_ADD3_D1B},
    ndkeywordstyle=\color{red},
    identifierstyle=\color{black},
    sensitive=false,
    comment=[l]{//},
    commentstyle=\color{gray}\ttfamily,
    stringstyle=\color{green}\ttfamily,
    morestring=[b]',
    morestring=[b]"
}

\def\BibTeX{{\rm B\kern-.05em{\sc i\kern-.025em b}\kern-.08em
    T\kern-.1667em\lower.7ex\hbox{E}\kern-.125emX}}
\begin{document}

\title{Automating Code Generation for Semiconductor Equipment Control from Developer Utterances with LLMs}
\author{
\begin{tabular}{ccc}
    \begin{tabular}{@{}c@{}}
      Youngkyoung Kim\textsuperscript{*} \\
      \textit{Department of Electrical} \\
      \textit{and Computer Engineering} \\
      Sungkyunkwan University \\
      Republic of Korea \\
      agnes66@skku.edu
    \end{tabular} &
    \begin{tabular}{@{}c@{}}
      Sanghyeok Park\textsuperscript{*} \\
      \textit{Department of Semiconductor} \\
      \textit{Display Engineering} \\
      Sungkyunkwan University \\
      \textit{Samsung Institute of Technology} \\
      \textit{Samsung Electronics} \\
      Republic of Korea \\
      sh88park@g.skku.edu
    \end{tabular} &
    \begin{tabular}{@{}c@{}}
      Misoo Kim \\
      \textit{Department of Artificial} \\
      \textit{Intelligence Convergence} \\
      Chonnam National University \\
      Republic of Korea \\
      misoo.kim@jnu.ac.kr
    \end{tabular}
    \\[1em]
    \begin{tabular}{@{}c@{}}
      Gangho Yoon \\
      \textit{Department of Semiconductor} \\
      \textit{Display Engineering} \\
      Sungkyunkwan University \\
      Republic of Korea \\
      yunkh21@g.skku.edu
    \end{tabular} &
    \begin{tabular}{@{}c@{}}
      Eunseok Lee\textsuperscript{†} \\
      \textit{College of Computing and Informatics} \\
      Sungkyunkwan University \\
      Republic of Korea \\
      leees@skku.edu
    \end{tabular} &
    \begin{tabular}{@{}c@{}}
      Simon S. Woo\textsuperscript{†} \\
      \textit{College of Computing and Informatics} \\
      Sungkyunkwan University \\
      Republic of Korea \\
      swoo@g.skku.edu
    \end{tabular}
\end{tabular}
}

\maketitle

\footnotetext[1]{\textsuperscript{*}These authors contributed equally to this work (co-first authors).}
\footnotetext[2]{\textsuperscript{†}These authors contributed equally as corresponding authors.}

\begin{abstract}
Semiconductors form the backbone of modern electronics, with their manufacturing and testing relying on highly specialized equipment and domain-specific programming languages. Equipment languages such as the Algorithmic Pattern Generator (ALPG) are critical for precise hardware control but are challenging to program due to their low-level syntax and steep learning curve. While large language models (LLMs) have shown promise in generating high-level code from natural language, their effectiveness on low-level equipment languages remains limited. To address this, we propose Progressive Knowledge Enhancement (PKE), a novel multi-stage prompting framework that progressively extracts and activates the latent knowledge within LLMs, guiding them from simple to complex examples without extensive fine-tuning. Empirical evaluation on an industrial ALPG dataset shows that PKE significantly outperforms standard prompting and surpasses state-of-the-art methods in generating correct ALPG code, achieving 11.1\% and 15.2\% higher exact match scores compared to the second-best technique. Further analysis of individual components confirms that progressive knowledge extraction based on difficulty enhances accuracy. Our study offer a practical approach to boosting LLM capabilities for specialized low-level programming, supporting greater productivity in semiconductor software development.

\end{abstract}


\section{Introduction}
\label{sec:introduction}


As semiconductors become increasingly important, the complexity of equipment control has also grown. Equipment control languages, such as ALPG \cite{lee2024new, lee2024cost}, are specialized for direct hardware communication and precise device configuration. Unlike general-purpose languages, they emphasize low-level, equipment-specific instructions, including nanosecond-level timing, custom commands, and domain-specific syntax for tasks like chip enable or write enable. Their unique features—such as specialized timing commands (e.g., \texttt{STPS TS1}), direct hardware manipulation, and unconventional control flow constructs (e.g., \texttt{MODULE}, \texttt{JSR})—distinguish them from mainstream languages and present significant challenges for new developers. Although essential for developing equipment control systems, the translation of natural language utterances \cite{mordechai2024novicode} into specialized control languages remains a manual and error-prone process that can hinder productivity. Programming these languages requires mastery of both non-human friendly syntax and hardware-specific knowledge \cite{matai2014enabling}, creating a steep learning curve. The scarcity of public resources and documentation further compounds these difficulties, as domain expertise is limited and support communities are small compared to mainstream programming languages.

Recent advances in Large Language Models (LLMs) have improved development productivity\cite{peng2023impact, hou2024large, cui2024productivity} by automating code generation from natural language descriptions~\cite{chen2021evaluating, jiang2024survey}. LLM-based code generation has demonstrated remarkable success across various high-level programming languages such as Python, Java, and JavaScript, where abundant training data and comprehensive documentation enable effective learning~\cite{brown2020language, wang2021codet5, li2023starcoder, roziere2023code}. Models like Codex~\cite{chen2021evaluating}, StarCoder~\cite{li2023starcoder}, and CodeGen~\cite{nijkamp2022codegen} have achieved impressive performance on established benchmarks like HumanEval~\cite{chen2021evaluating} and MBPP~\cite{austin2021program}, demonstrating their capability to understand high-level human intent and translate it into executable code.

However, the effectiveness of existing NL2Code approaches is greatly reduced in specialized, low-resource domains like ALPG, particularly due to the proprietary nature that result in limited public availability of such languages \cite{ciborowska2021contemporary, joel2024survey}. While some progress has been made for high-level hardware description languages like Verilog~\cite{thakur2024verigen}, low-level equipment control languages like ALPG present an even greater challenge, as LLMs have minimal exposure to their unique syntax and operational constraints during training~\cite{brown2020language,wei2022chain}. As a result, standard prompting techniques, including few-shot learning and Chain-of-Thought (CoT) reasoning, fail to properly elicit specialized control semantics. This limitation is illustrated in Table~\ref{tab:example}, where the LLM partially recognizes the intent to test flash memory but cannot generate the correct commands. 

Nevertheless, the model’s generated explanation reveals a nuanced understanding of the NAND Flash command sequence. Although the produced code does not fully conform to the required ALPG syntax, the model demonstrates strong analytical capability in delineating the command, address, and data phases. Furthermore, the model’s recognition of the need for a mnemonic language highlights its underlying grasp of hardware concepts and low-level programming constructs. We hypothesize that LLMs possess latent understanding of low-level concepts crucial for hardware control acquired in training lines. The key challenge, however, is that this knowledge is often not activated during inference. 

\renewcommand{\arraystretch}{1.0}
\begin{table}[t]
\centering
\caption{Motivating Example of developer utterance and answer code and gpt-4.1 output with CoT prompting}
\label{tab:example}
\resizebox{\columnwidth}{!}{%
\begin{tabular}{cl}
\hline
Utterance & \begin{tabular}[c]{@{}l@{}}Write an ALPG pattern program code\\ that performs the 85h Command -\\ Address 5cycle - Data in.\end{tabular} \\ \hline
\begin{tabular}[c]{@{}c@{}}Answer\\ Code\end{tabular} & \begin{tabular}[c]{@{}l@{}}JSR G\_LF001\_CMDI               CE0  TP\textless{}\#85       TS2\\ JSR G\_LF001\_ADD5\_D1\_D2          CE0               TS1\\ STPS  TS1\end{tabular} \\ \hline
\begin{tabular}[c]{@{}c@{}}Gpt-4.1\\ Generated\\ Code\\ with\\ CoT\end{tabular} & \begin{tabular}[c]{@{}l@{}}Absolutely, let's break this down step by step. \\ The  85h Command - Address 5cycle - Data in \\ pattern is common in NAND Flash memory interfaces. \\ ALPG (Algorithm Pattern Generator) programs are \\ used in testers to simulate such command sequences to devices.\\ (ommitted additional descriptions)\\ \\ [-5pt]; -- Initialization (if needed)\\ RESET       ; Optional: Reset device to known state\\ \\[-3pt] ; -- Command Phase\\ CMD\_WRITE   85h         ; Send 0x85 Command on the command bus\\ \\[-3pt] ; -- Address Phase (5 Cycles)\\ ADDR\_WRITE  addr0       ; 1st address cycle (A0-A7)\\ ADDR\_WRITE  addr1       ; 2nd address cycle (A8-A15)\\ ADDR\_WRITE  addr2       ; 3rd address cycle (A16-A23)\\ ADDR\_WRITE  addr3       ; 4th address cycle (A24-A31)\\ ADDR\_WRITE  addr4       ; 5th address cycle (A32-A39)\\ \\[-3pt] ; -- Data-In Phase\\ DATA\_WRITE  data0       ; 1st data byte/word in\\ DATA\_WRITE  data1       ; 2nd data byte/word in\\ ; ... repeat for as many data cycles as needed\end{tabular} \\ \hline
\end{tabular}%
}
\vspace{-15pt}  

\end{table}



To address this limitation, we draw inspiration from human learning processes—specifically, the practice of gradually building knowledge from simple to complex examples, as advocated in curriculum learning \cite{bengio2009curriculum}. We propose \textbf{Progressive Knowledge Enhancement (PKE)}, which systematically elicits and applies the model’s latent expertise to improve code generation accuracy. PKE guides the model toward domain alignment by gradually constructing background knowledge through easy to hard examples and leveraging this knowledge during code generation. Our framework comprises three core components: (1) LLM-based difficulty annotation, (2) progressive background knowledge construction from easy to hard examples, and (3) guided code generation using the enhanced knowledge. This enables LLMs to incrementally develop a deeper understanding of specialized equipment control languages, improving domain-specific reasoning and code quality.

Our empirical study, conducted on a real-world ALPG dataset curated from industrial use cases, demonstrates the effectiveness of PKE. The main contributions of this work are as follows. First, we introduce PKE, a novel prompting framework designed to improve LLM-based code generation in low-resource, specialized programming domains. Second, we present empirical results demonstrating that PKE significantly outperforms standard prompting baselines and adapted state-of-the-art (SOTA) methods, achieving notable gains in across all metrics on industrial ALPG code generation tasks. Our study establish a practical framework for applying LLMs to specialized semiconductor equipment control languages and demonstrate the viability of prompting-based solutions for industrial applications involving proprietary languages.

\begin{figure*}[t]
    \centering
    \includegraphics[width=0.9\textwidth]{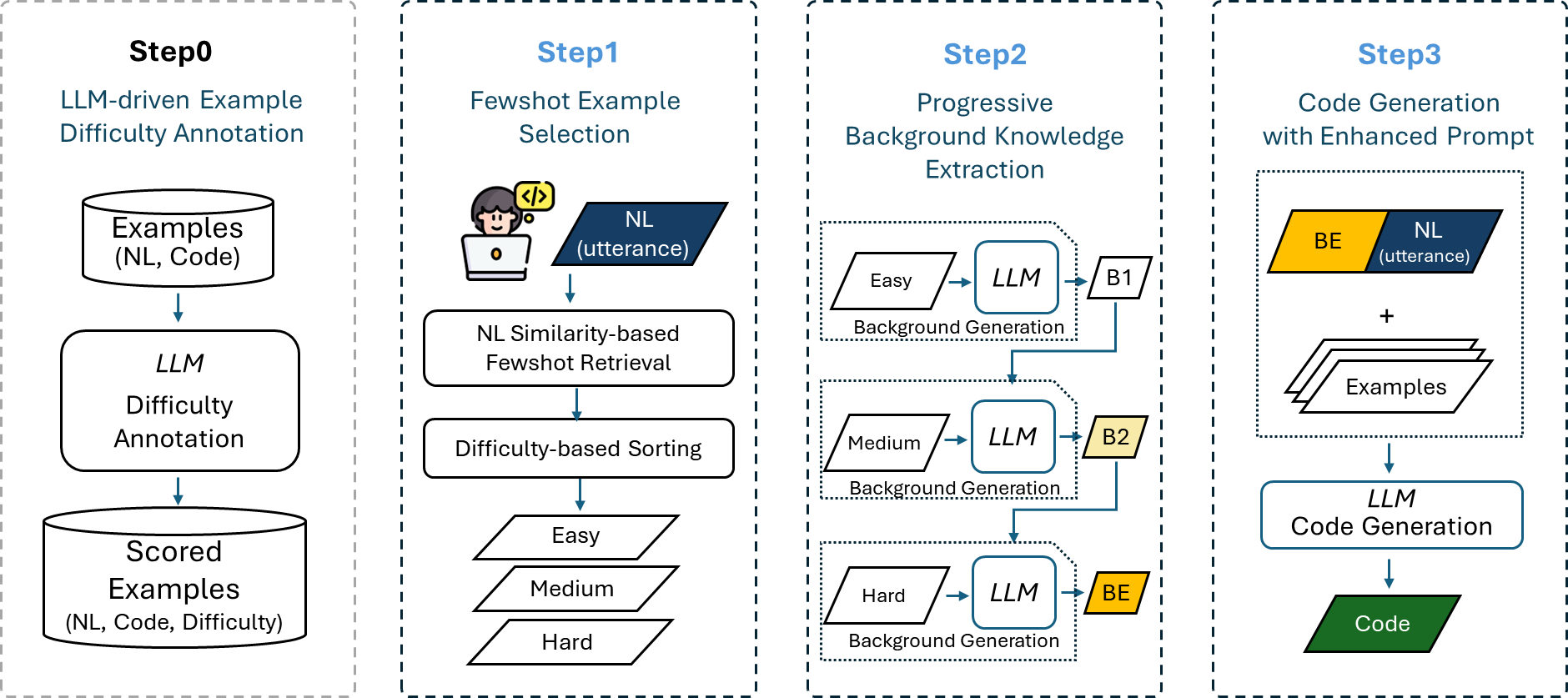} 
    \caption{PKE Framework Overview. The framework consists of four main stages: Step 0 performs LLM-adaptive difficulty annotation, Step 1 conducts dynamic few-shot example selection, Step 2 implements sequential reasoning enhancement, and Step 3 generates the final ALPG code.}
    \label{fig:framework}
\end{figure*}

\section{Proposed Method: PKE}
\label{sec:proposed_method}

We propose PKE, a novel framework designed to systematically extract embedded knowledge in LLMs utilizing examples progressively. Figure~\ref{fig:framework} illustrates the complete PKE workflow.

\subsection{Step 0: LLM-Adaptive Difficulty Annotation}
\label{subsec:step0_difficulty_annotation}

This step only needs to be performed once for the example corpora and does not need to be repeated for every new input. PKE begins by creating a difficulty-scored corpus of ALPG examples using LLM-based assessment. For each \texttt{(Natural Language, ALPG Code)} pair in the initial corpus, we employ an LLM to evaluate generation difficulty on a 0-100 scale based on multiple criteria as in Listing \ref{lst:difficulty}. These criteria provide general guidance to help the LLM assess the relative difficulty of ALPG code generation, considering the unique characteristics that make semiconductor control languages more challenging than general programming languages.

\begin{lstlisting}[style=textstyle, basicstyle=\scriptsize, caption={Difficulty annotation prompt template.}, label={lst:difficulty}]
Judge the difficulty (0-100) to generate the 
following ALPG code from the given natural 
language description:

Natural Language: {nl_description}
ALPG Code: {alpg_code}

Rate the difficulty from 0 (very easy) to 100 (very hard) based on:
- Syntax complexity
- Hardware interaction requirements  
- Timing constraints
- Overall implementation challenge

Difficulty Score: 
\end{lstlisting}

To ensure robustness, we repeat this scoring process five times per example with different temperature setting [0, 0.2, 0.4, 0.6, 0.8]. The average of the final scores is then used as the overall difficulty score for each example.

\subsection{Step 1: Few-Shot Example Selection}
\label{subsec:step1_dynamic_selection}

For each input query, PKE employs NL-similarity-based retrieval to select representative examples from the example pool. The framework uses similarity metrics (such as BM25~\cite{robertson2009probabilistic}) to identify semantically relevant examples, which are then sorted by difficulty score. This selection process provides the LLM with examples that span different difficulty levels, which will be further enhanced with background knowledge in the subsequent step.

\subsection{Step 2: Sequential Reasoning Enhancement}
\label{subsec:step2_reasoning_enhancement}

This stage progressively builds domain-specific background knowledge through three sequential steps, each building upon the previous:

\begin{lstlisting}[style=textstyle, basicstyle=\scriptsize, caption={Sequential reasoning enhancement prompt template.}, label={lst:sre_prompts}]
### Code description: 
{code_description}

### Corresponding Code: 
{code}

{previous_knowledge}

Please write necessary domain knowledge to 
help generate the corresponding ALPG code 
based on the code description. The goal is to 
help a language model better understand and 
generate the corresponding ALPG code.
\end{lstlisting}

The LLM sequentially processes Easy, Medium, and Hard examples to build its background knowledge. It begins by analyzing the Easy example to attain background knowledge ($B_1$), then expands this knowledge with the Medium-difficulty example by ($B_2$). Finally, using the accumulated knowledge, the LLM examines the Hard example to produce final background knowledge ($B_E$). This sequential approach mirrors human learning progression from simple to complex examples, enabling systematic knowledge accumulation and refinement.

\subsection{Step 3: Enhanced Code Generation}
\label{subsec:step3_final_generation}

PKE constructs the final prompt by concatenating: (1) the input NL query  (including any additional specifications or constraints), (2) the selected few-shot examples, and (3) the enhanced background knowledge $B_E$ as in Listing \ref{lst:final}. The enhanced background knowledge $B_E$ serves as explicit guidance for understanding ALPG programming domain, while the few-shot examples provide concrete implementation patterns. Placing $B_E$ first allows the LLM to internalize domain principles before examining concrete examples, enabling better understanding of the reasoning behind each implementation pattern. This rich and more refined context enables the LLM to generate more accurate and contextually appropriate ALPG code by leveraging both enhanced domain comprehension.

\begin{lstlisting}[style=textstyle, basicstyle=\scriptsize, caption={Final code generation prompt template.}, label={lst:final}]
You are an expert in {lang} programming. 

Enhanced Background Knowledge (generated from progressive difficulty examples):
{enhanced_background_knowledge}

Relevant Few-shot Examples:
{fewshot_examples}

Using the above knowledge and examples as guidance, generate accurate {lang} code for:

### Code description
{Utterance}
### Corresponding Code
\end{lstlisting}

\section{Experimental Setup}

\textbf{Experimental Dataset.}
\label{subsec:alpg_dataset}
Our dataset comprises 271 NL-ALPG code pairs for semiconductor test equipment, curated from developer requests. Ground-truth ALPG code was expert-verified for accuracy and compliance. Due to proprietary and security constraints, the dataset is not publicly available. 

\textbf{LLM Model. }
We utilized in-house LLMs with approximately 70B and 65B parameters respectively for all our experiments. We selected these models to ensure data privacy and appropriate security measures for proprietary data handling.

\textbf{Evaluation Metrics. }
To evaluate ALPG code generation, we use Exact Match (EM), BLEU, and Levenshtein distance. EM reports the percentage of generated code snippets that exactly match the ground truth after normalization~\cite{chen2021evaluating,austin2021program}. BLEU~\cite{papineni2002bleu} measures n-gram overlap between generated and reference code and is widely used in code generation evaluation~\cite{yin2017syntactic,rabinovich2017abstract}. Levenshtein distance~\cite{chen2024survey} quantifies the minimum word-level edits needed to transform the generated code into the reference, reflecting post-editing effort.

\subsection{Baselines}

We compare PKE with several baseline prompting strategies, including state-of-the-art code generation techniques. The baselines are: (1) \textbf{Zero-shot Prompting}, where the LLM generates ALPG code from the input NL query without examples; (2) \textbf{Standard Few-shot Prompting (Similarity-based)}, which provides the LLM with a few ALPG examples selected via BM25 retrieval~\cite{robertson2009probabilistic} based on NL similarity to the input; (3) \textbf{MapCoder}~\cite{islam2024mapcoder}, a multi-agent approach emulating example recall, planning, code generation, and debugging; and (4) \textbf{$\mu$Fix}~\cite{tian2025fixing}, which iteratively refines the LLM’s understanding of code and specifications. We minimally adapted SoTA techniques by adding few-shot examples to restrict outputs to our target language and modifying prompts (e.g., replacing “Java” with “ALPG”) to ensure alignment with our specifications.

\subsection{Research Questions (RQs)}
\label{subsec:research_questions}
Our experiments are designed to answer the following research questions:
\begin{itemize}
    \item \textbf{RQ1.} To what extent does the proposed framework improve ALPG code generation performance over established baselines?

    \item \textbf{RQ2.} How does the \textit{progressive} knowledge extraction influence code generation accuracy in PKE?
    
    \item \textbf{RQ3.} How sensitive are LLM to difficulty of examples in PKE framework?
        

\end{itemize}

\section{Results}
\label{sec:results}

\subsection{RQ1. Comparison to Prompting Strategies}
\label{subsec:rq1_overall_performance}

\renewcommand{\arraystretch}{1.2}
\begin{table}[t]
\centering

\caption{Comparison of Different Prompting Strategies for ALPG Code Generation. \textcolor{Blue}{Ratio} shows improvements over the second-best approach, few-shot.}
\label{tab:main_performance_comparison}
\resizebox{\columnwidth}{!}{%
\begin{tabular}{cccc|ccc}
\hline
 & \multicolumn{3}{c|}{Model-70B} & \multicolumn{3}{c}{Model-65B} \\
 & EM & BLEU & Leven & EM & BLEU & Leven \\ \hline
Zero-shot & 0.016 & 0.085 & 0.016 & 0.018 & 0.092 & 0.018 \\
Few-shot & 0.283 & 0.718 & 0.782 & 0.289 & 0.651 & 0.719 \\ \hline
MapCoder & 0.032 & 0.152 & 0.032 & 0.028 & 0.150 & 0.030 \\
$\mu$Fix & 0.000 & 0.118 & 0.000 & 0.000 & 0.105 & 0.000 \\ \hline
PKE & \textbf{0.315} & \textbf{0.745} & \textbf{0.813} & \textbf{0.333} & \textbf{0.728} & \textbf{0.790} \\[-0.12cm]
 & \scriptsize\textcolor{Blue}{+11.1\%} & \scriptsize\textcolor{Blue}{+3.8\%} & \scriptsize\textcolor{Blue}{+4.0\%} & \scriptsize\textcolor{Blue}{+15.2\%} & \scriptsize\textcolor{Blue}{+11.8\%} & \scriptsize\textcolor{Blue}{+9.9\%} \\ \hline
\end{tabular}%
}
\end{table}

Table~\ref{tab:main_performance_comparison} shows that our PKE framework outperformed all other prompting strategies across metrics. In the zero-shot setting, models performed poorly due to lack of familiarity with relevant programming keywords. The few-shot approach, using similar examples, improved results (EM 0.283, BLEU 0.718, Leven 0.782), but substantial performance gaps remained, underscoring the need for better domain alignment. Notably, SoTA techniques still struggled with our target language, often generating only superficially correct or inaccurate ALPG code. For instance, $\mu$Fix's specification analysis produced shallow output, irrelevant to domain-specific details, generating code in wrong syntax.


Incorporating background knowledge through the proposed PKE framework yielded substantial performance gains. Specifically, EM accuracy increased by 11\% and 15\% for the 70B and 65B parameter models, respectively. Similarly, BLEU scores improved by 4\% and 12\%, while Levenshtein distance exhibited improvements of 4\% and 10\% for the respective models. These results indicate that enriching input prompts with background knowledge significantly reduces ambiguity in code generation, leading to outputs that require fewer edits and better align with the intended functionality.

\subsection{RQ2. Impact of progressive knowledge extraction}
\label{subsec:rq3}

\begin{table}[t]
\centering
\caption{Extraction ordering basis variation: retrieval similarity (Sim) vs. LLM-annotated difficulty (PKE).}
\label{tab:rq3.1}
\large
\resizebox{\columnwidth}{!}{%
\begin{tabular}{ccc|cc}
\hline
 &  & Without BE & \multicolumn{2}{c}{With Extracted BE} \\
 &  & (Few Shot) & Sim & PKE \\ \hline
\multirow{3}{*}{Model-70B} & EM & 0.283 & 0.289 (+2.2\%) & \textbf{0.315 (+11.1\%)} \\
 & BLEU & 0.718 & 0.721 (+0.4\%) & \textbf{0.745 (+3.7\%)} \\
 & Leven & 0.782 & 0.786 (+0.5\%) & \textbf{0.813 (+3.9\%)} \\ \hline
\multirow{3}{*}{\begin{tabular}[c]{@{}c@{}}Model-65B\end{tabular}} & EM & 0.289 & 0.315 (+8.7\%) & \textbf{0.333 (+15.2\%)} \\
 & BLEU & 0.651 & 0.716 (+9.9\%) & \textbf{0.728 (+11.7\%)} \\
 & Leven & 0.719 & 0.788 (+9.6\%) & \textbf{0.790 (+9.8\%)} \\ \hline
\end{tabular}%
}
\end{table}
    
We analyze our core design choice: performing knowledge extraction in a \textit{progressive} manner. To validate this, we compared the baseline few-shot setting (without background knowledge) to a progressive extraction strategy that orders examples by similarity (Sim). As shown in Table \ref{tab:rq3.1}, progressive extraction based on similarity yields consistent improvements over the few-shot baseline across all metrics. For example, Exact Match (EM) improves from 0.283 to 0.289 (+2.2\%) for the 70B parameter model and from 0.289 to 0.315 (+8.7\%) for the 65B parameter model. Similarly, BLEU and Levenshtein also show noticeable gains, particularly for the 65B parameter model, where BLEU increases by 9.9\% and Levenshtein by 9.6\%. While progressive extraction based on similarity improves performance, introducing difficulty awareness by ordering knowledge from easy to hard yields even greater accuracy. PKE, which integrates both progressive extraction and difficulty awareness, consistently outperforms the similarity-order enhancement (Sim) across all metrics and both models, improving EM by 8.7\% and 6.0\%, and surpasses few-shot learning by 11.1\% and 15.2\%, respectively. These findings highlight the synergistic effect of combining progressive extraction with difficulty-aware ordering, as this integration enhances the model's ability to leverage background knowledge and focus on the target language, underscoring the critical role of extraction order.

\subsection{RQ3. Sensitivity to example difficulty}
\label{subsec:rq2}

\begin{figure}
    \centering
\includegraphics[width=0.9\columnwidth]{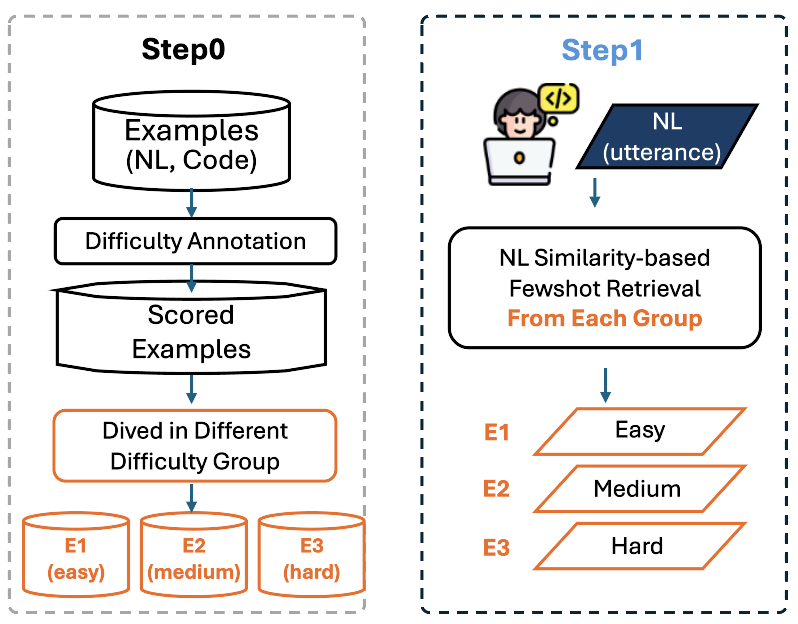}
    \caption{D\&R Configuration in Step0 and step1. }
    \label{fig:dnr}
\end{figure}


\textbf{Configuration:} To assess the impact of example difficulty, we designed a Divide-and-Retrieve (D\&R) setup (Figure~\ref{fig:dnr}), which modifies PKE by partitioning the example corpus into easy, medium, and hard groups in step 0. Modifications to PKE are highlighted in orange. D\&R retrieves examples from specified difficulty levels, enabling evaluation of various configurations: EMH (each from easy, medium, and hard), as well as single-difficulty settings (EEE, MMM, HHH), where all retrieved examples come from the same group. 

\renewcommand{\arraystretch}{1.0}
\begin{table}[t]
\centering
\normalsize
\caption{Comparison with Divide-and-Retrieve (D\&R). The best overall performance is highlighted in bold, while the highest performance among the D\&R strategies is indicated with underline.}
\label{tab:rq2}
\resizebox{\columnwidth}{!}{%
\begin{tabular}{lc|ccc|ccc}
\hline
 & \multicolumn{1}{l|}{} & \multicolumn{3}{c|}{{\color[HTML]{242424} Model-70B}} & \multicolumn{3}{c}{{\color[HTML]{242424} Model-65B}} \\
 & \multicolumn{1}{l|}{} & EM & BLEU & Leven & EM & BLEU & Leven \\ \hline
 & FS & 0.283 & 0.718 & 0.782 & 0.289 & 0.651 & 0.719 \\
 & PKE & \textbf{0.315} & \textbf{0.745} & \textbf{0.813} & \textbf{0.333} & \textbf{0.728} & \textbf{0.790} \\ \hline
\multicolumn{1}{c}{} & EMH & 0.258 & \uline{0.726} & \uline{0.795} & \uline{0.226} & \uline{0.646} & \uline{0.725} \\
\multicolumn{1}{c}{} & EEE & \uline{0.289} & 0.525 & 0.701 & 0.201 & 0.453 & 0.536 \\
\multicolumn{1}{c}{} & MMM & 0.013 & 0.369 & 0.467 & 0.006 & 0.272 & 0.366 \\
\multicolumn{1}{c}{\multirow{-4}{*}{\begin{tabular}[c]{@{}c@{}}D\\ \&\\ R\end{tabular}}} & HHH & 0.000 & 0.316 & 0.392 & 0.000 & 0.268 & 0.350 \\ \hline
\end{tabular}%
}
\end{table}

\textbf{Sensitivity to Difficulty:} 
We found that model performance declined sharply as example difficulty increased: EM scores for the 70B parameter model dropped from 0.289 (EEE) to 0.013 (MMM) to 0.000 (HHH), and for the 65B parameter model from 0.201 to 0.006 to 0.000. However, arranging examples from different levels (EMH) in a diverse manner yielded the best performance among the D\&R variants, underscoring the benefit of gradual knowledge extraction.

\textbf{Sensitivity to Difficulty Diversity:} 
Comparing EMH and PKE, PKE consistently outperformed EMH across all metrics as shown in Table~\ref{tab:rq2}. For the 70B parameter model, PKE achieved EM 0.315, BLEU 0.745, and Leven 0.813, notably surpassing EMH (EM 0.226, BLEU 0.646), improving EM by 39\%. The 65B parameter model showed similar trends. These results indicate that while difficulty progression aids learning, prioritizing example similarity is more effective, supporting our retrieve-and-sort approach.

\section{Related Work}
\label{sec:related_work}

LLMs excel at NL2Code generation in high-level languages (e.g., Python, Java) thanks to ample data and prompting methods like few-shot and CoT ~\cite{jiang2024survey, brown2020language, chen2021evaluating, wei2022chain}. However, their effectiveness drops in specialized, low-resource domains. PKE addresses this gap by progressively enhancing domain-specific knowledge through tailored procedure.  Models like Codex, StarCoder, and CodeGen have demonstrated strong performance on benchmarks such as HumanEval and MBPP~\cite{chen2021evaluating, li2023starcoder, nijkamp2023codegen, austin2021program}.
LLM use in hardware programming mainly targets HDLs such as Verilog, but ALPG and similar control languages present unique challenges due to lower abstraction, resource scarcity, and hardware-specific dependencies~\cite{blocklove2023chip, thakur2024verigen}. Existing transfer or fine-tuning methods ~\cite{cassano2023knowledge, chen2022transferability} often assume more data than available for proprietary languages. Prompting-based approaches, like PKE, offer a practical solution by improving LLM performance without requiring large, domain-specific datasets~\cite{liu2023opportunities}.


\section{Conclusion}
\label{sec:conclusion}

This work introduces Progressive Knowledge Extraction (PKE), a method that attempt to extract the knowledge embedded in LLMs by guiding them from easy to difficult examples in a progressive manner. Through validation on the ALPG dataset, a representative industrial equipment programming language used for semiconductor, we show that PKE significantly outperforms standard prompting baselines and surpasses SOTA methods in generating correct ALPG code. Our studies on design choices further confirm that progressive background knowledge extraction plays a critical role in improving generation accuracy. Our study provide a practical and effective approach to enhancing LLM capabilities for specialized low-level hardware control programming, offering a pathway to increased automation and productivity in equipment software development. In future work, we will investigate the extent of actual productivity improvement and examine whether model-extracted knowledge complements human-provided knowledge.

\bibliographystyle{./IEEEtran} 
\bibliography{./IEEEabrv,./IEEEexample} 

\end{document}